\documentstyle[11pt]{article}
\begin{document}

\title{Rigid open membrane and non-abelian non-commutative Chern-Simons theory }

\author{Yi-Xin Chen\thanks{Email:yxchen@zimp.zju.edu.cn} 
\\[.3cm]
Zhejiang Institute of Modern Physics, Zhejiang University,\\ 
             Hangzhou 310027, P. R. China
}

\date{\today}
\maketitle

\begin{abstract}

In the Berkooz-Douglas matrix model of M theory in the presence of longitudinal $M5$-brane, we investigate the effective dynamics of the system by considering the longitudinal $M5$-brane as the background and the spherical $M5$-brane related with the other space dimensions as the probe brane. Due to there exists the background field strength provided by the source of the longitudinal $M5$-brane, an open membrane should be ended on the spherical $M5$-brane based on the topological reason. The formation of the bound brane configuration for the open membrane ending on the 5-branes in the background of longitudinal 5-brane can be used to model the 4-dimensional quantum Hall system proposed recently by Zhang and Hu. The description of the excitations of the quantum Hall soliton brane configuration is established by investigating the fluctuations of $D0$-branes living on the bound brane around their classical solution derived by the transformations of area preserving diffeomorphisms of the open membrane. We find that this effective field theory for the fluctuations is an $SO(4)$ non-commutative Chern-Simons field theory. The matrix regularized version of this effective field theory is given in order to allow the finite $D0$-branes to live on the bound brane. We also discuss some possible applications of our results to the related topics in M-theory and to the 4-dimensional quantum Hall system. 

\indent

\vspace{.5cm}

\end{abstract}

\setcounter{equation}{0}

\section{Introduction}

\indent

A configuration of branes and strings, called as the quantum Hall soliton, was discussed with low energy dynamics similar to those of condensed matter system displaying the fractional quantum Hall effect\cite{Laughlin} in \cite{Bernevig}. The configuration consists of a spherical $D2$-brane wrapping some flat $D6$-branes, in which the same fundamental strings as the number of $D6$-branes must stretch from the $D6$-branes at the center to the spherical $D2$-brane according to the consideration of the configuration topology. The string ends on the $D2$-brane play the role of the electrons. The magnetic flux quanta are $N$ $D0$-branes dissolved in the $D2$-brane. In this way, several phenomena that occur in 2-dimensional quantum Hall system can be modeled qualitatively in terms of the strings and branes involved in the configuration. The authors in \cite{Bernevig} also showed how to describe the background magnetic field in terms of an incompressible fluid of $D0$-branes using Matrix theory. The fluctuations of the $D0$-branes around the classical solution, which are regarded as the dynamical degrees of freedom of the excitations in the quantum Hall soliton configuration, are described by the $U(1)$ non-commutative Chern-Simons theory.

Another line of progress was started by the proposal of Susskind \cite{Susskind} that the granular structure of the quantum Hall fluid can be captured by making the ordinary Chern-Simons description non-commutative. He proposed that the $U(1)$ non-commutative Chern-Simons theory on the plane may provide a description of the (fractional) quantum Hall fluid, and specifically of the Laughlin states. Susskind's non-commutative 
Chern-Simons theory on the plane describes a spatially infinite quantum 
Hall system. It, i.e., does the Laughlin states at filling fractions $\nu$ 
for a system of an infinite number of electrons confined in the lowest 
Landau level. The fields of this theory are infinite matrices that act on 
an infinite Hilbert space, appropriate to account for an infinite number of 
electrons. Subsequently, Polychronakos \cite{Polychronakos} proposed a 
matrix regularized version of Susskind's non-commutative Chern-Simons 
theory in an effort to describe finite systems with a finite number of 
electrons in the limited spatial extent. This matrix model was shown to 
reproduce the basic properties of the quantum Hall droplets and two special 
types of excitations of them. Furthermore, it was shown that there 
exists a complete minimal basis of exact wave functions for the matrix 
regularized version of non-commutative Chern-Simons theory at arbitrary 
level $\nu^{-1}$ and rank $N$, and that those are one to one 
correspondence with Laughlin wave functions describing excitations of a 
quantum Hall droplet composed of $N$ electrons at filling fraction $\nu$ 
\cite{Hellerman}. It is believed that the matrix regularized version of 
non-commutative Chern-Simons theory is precisely equivalent to the theory 
of composite fermions in the lowest Landau level, and should provide an 
accurate description of fractional quantum Hall state. However, it does 
appear an interesting conclusion that they are agreement on the long 
distance behavior, but the short distance behavior is different 
\cite{Karabali}.

Recently, Zhang and Hu \cite{Zhang}, on the four-sphere $S^4$, have found a
4-dimensional generalization of the quantum Hall system that is composed of 
many particles moving in four dimensional space under a $SU(2)$ gauge 
field. Since at special filling factors, the quantum disordered ground 
state of 4-dimensional quantum Hall effect is separated from all excited 
states by a finite energy gap, the lowest energy excitations are 
quasi-particle or quasi-hole excitations near the lowest Landau level. The 
quantum disordered ground state of 4-dimensional quantum Hall effect is 
the state composed coupling by particles lying in the lowest Landau level 
state. At appropriate integer and fractional filling fractions, the system 
forms an incompressible quantum liquid. The authors in the references \cite{Hu,Fabinger,Chen,Karabali1,Kimura,Bernevig1,Ramgoolam,Elvang,Chen1,Chen2,Zhang1} have developed the idea of Zhang and Hu in other directions, especially, and have explored some non-commutative structures related with the 4-dimensional quantum Hall system proposed by Zhang and Hu.

Based on the 2-dimensional quantum Hall fluid is very closely related to the $U(1)$ non-commutative Chern-Simons theory, alternative brane constructions for the 2-d quantum Hall system in string theory had appeared in the recent papers\cite{Brodie,Gubser,Hellerman1,Freivogel}. However, these papers did not involved any brane construction for the higher dimensional quantum Hall systems. But, for Zhang and Hu's 4-dimensional quantum Hall system, a string realization based on the brane construction was established by Fabinger\cite{Fabinger}. The quantum Hall soliton configuration proposed by Fabinger is composed of a $D4$-brane, a stack of coincident spherical $D4$-branes and the fundamental strings stretching between them. The four-sphere $S^4$, which is the space of the particle's living of the 4-dimensional quantum Hall system, is modeled by this stack of coincident spherical $D4$-branes with a homogeneous instanton in their world volume. The charged particles of the Hall system themselves can be modeled as the ends of fundamental strings connecting the spherical $D4$-branes to the flat $D4$-brane placed at the center of the $S^4$. 

It is well known that in the brane constructions for the quantum Hall system in string theory, an essential ingredient is the $U(1)$ non-commutative Chern-Simons theory effectively describing the fluctuations of the quantum Hall soliton configuration. What is the effective field theory describing the fluctuations of the quantum Hall soliton configuration based on the brane construction for Zhang and Hu's 4-dimensional quantum Hall effect? The goal of this paper is to answer this problem.

In order to investigate the above problem, we shall revisit the Berkooz and Douglas's matrix model\cite{Berkooz} which the modified version of the BFSS model\cite{Banks} in M theory. If the longitudinal 5-brane is considered as the background at the center of a spherical 5-brane taken as the probe brane, we find that this background can be effectively described by an Yang's $SU(2)$ monopole\cite{Yang} at the center. Based this fact, we construct the brane realization of the Zhang and Hu's 4-dimensional quantum Hall fluid given by the quantum Hall soliton configuration $M5_c-M2-M5_s$, where the subscripts $c$ and $s$ stand for the longitudinal 5-brane at the center and the spherical 5-brane respectively. The open membrane $M2$ ending on the spherical 5-brane consists of the bound brane $M2-M5_s$ in which there are $N$ $D0$-branes dissolved. The fluctuations of this quantum Hall soliton configuration are described by an $SO(4)$ non-commutative Chern-Simons field theory derived by the transformations of area preserving diffeomorphisms on the open membrane.

This paper is organized as follows. Section two introduces the 
Berkooz-Douglas matrix model as the starting point of our discussion. By considering the longitudinal 5-brane as the background and the spherical 5-brane composed of the transverse  space dimensions as the probe, we analyze the non-abelian Berry connection contributed from the zero modes of the fermion fields. This contribution can be effectively produced by Yang's $SU(2)$ monopole at the center of the spherical 5-brane. Section 
three includes the discussions that it is necessary that the open membrane must be ended on the spherical 5-brane to form the bound brane configuration in the four-form field strength created by the longitudinal 5-brane at the center, and the description of the bound brane configuration with the $D0$-branes. In section four, we find the effective field theory to describe the low energy dynamical degrees of freedom for the excitations of the bound brane around the classical solution by considering the transformations of area preserving diffeomorphisms on the open membrane. This effective field theory is an $SO(4)$ non-commutative Chern-Simons theory. Section five gives the non-commutative structure of the $S^4$, i.e., the spherical 5-brane, by performing the second Hopf mapping from the $S^7$ to $S^4$. We get the result in agreement with that given by Zhang and Hu. The matrix regularized version of the $SO(4)$ non-commutative field theory are also established in this 
section. The last section includes the summary of the results and remarks on some physical applications of them.

\section{The Berkooz-Douglas matrix model and the Berry connection induced by five-brane}

\indent

In this section, we shall review the matrix model of Berkooz and Douglas, and discuss some useful properties of this model. The contribution of the non-abelian Berry phase will be calculated in the presence of longitudinal 5-brane.

It is well known that branes can often have boundaries on higher dimensional brane \cite{Strominger}. Perhaps a simple example is that a $Dp$-brane can end on $D(p+2)$-brane, which is related by S and T-duality to the fact that fundamental strings may end on $D$-branes. In general, one can find the configurations of open $D(p+2)$-branes ending on $D(p+4)$-branes. The matrix model of Berkooz and Douglas \cite{Berkooz}, proposed to describe M-theory in the presence of longitudinal $M5$-branes, can provide such configurations of bound branes. This model should include states corresponding to arbitrarily shaped open membranes ending on the $M5$-branes, since these open membranes are allowed objects in M-theory. Interpreted in the context of type IIA string theory at low energies and weak coupling, these configurations should correspond to open $D2$-branes ending on the $D4$-branes\cite{Raamsdonk}. The Berkooz-Douglas matrix model was proposed as a matrix model for DLCQ M-theory with $N$ units of momentum along the longitudinal direction in the presence of $k$ longitudinal $M5$-branes. It is given by the limit of type IIA string theory with $N$ $D0$-branes and $k$ $D4$-branes.

The Berkooz-Douglas model is a modification of the M theory Lagrangian of BFSS model\cite{Banks} to describe a longitudinal five-brane. The Lagrangian of Berkooz-Douglas model \cite{Berkooz} is read as
\begin{eqnarray}
{\cal L} & = & Tr[\frac{1}{2R}D_{t}X^i D_{t}X^i -{\bar \theta}\gamma_{-}D_t \theta -R{\bar \theta}\gamma_{-}\gamma_{i}[\theta,X^i]
-\frac{1}{4}R[X^i,X^j]^2] \nonumber\\
& + &T r[|D_t V^{\rho{\dot \rho}}|^2 +\chi D_t \chi -V_{\rho {\dot \rho}}(X^a-x^a)^2 V^{\rho {\dot \rho}}-\chi^{\dot \rho}_{\alpha}(X^a-x^a)\gamma^{\alpha\beta}_{a}\chi_{{\beta}{\dot \rho}} \nonumber\\
& - &V_{\rho{\dot \rho}}(\theta-\theta_0 )^{\rho}_{\alpha}\chi^{\alpha{\dot \rho}}-V_{\rho{\dot \rho}}[X^m,X^n]\sigma^{\rho\sigma}_{mn}V_{\sigma}^{\dot \rho}-|V|^4 ],
\end{eqnarray}
here, $x^a$ are diagonal matrices to describe the transverse positions of the $M5$-brane. The terms in the first line of the equation (1) are just the Lagrangian of BFSS matrix model. $X^m$ with $1\leq m\leq 4$ represent the coordinates of the 4-dimensional hyperplane in which the longitudinal 5-brane is embedded, and $X^a$ with $5\leq m\leq 9$ the transverse coordinates with respect to this hyperplane. The manifest $SO(9)$ symmetry in the nine transverse dimensions is broken to $SO(4)_{||}\times SO(5)_{\perp}$. $\rho$ and ${\dot \rho}$ index the two spinor representations of $SO(4)_{||}$, and $\alpha$ do a spinor of $SO(5)_{\perp}$. Adding the complex boson $V^{\rho {\dot \rho}}$ and the fermion $\chi^{\alpha{\dot \rho}}$ to the BFSS matrix model is to describe the five-brane in the M theory. 

As the above mention, the lightest modes such strings form the hypermultiplet in the vector representation of the zero-brane gauge symmetry group. Keeping only these lightest modes as additional M theory degrees of freedom leads to a theory with no new massless degrees of freedom, but a modified dynamics for the zero-branes and new bound states. One can produce the effective Lagrangian at the low energy by integrating out the fields $V$ and $\chi$. Furthermore, we can take the position of the five-brane to be $x^a =0$, and investigate the brane configurations in the background of this five-brane. Different with Berkooz and Douglas considering a membrane in this background, however, we shall study the more general configurations with respect to $X^a$ in this background. If one finishes the integration of the fermion $\chi$, the topologically non-trivial contribution will appear in the effective Lagrangian. Berkooz and Douglas had shown that the five-brane induces a Berry phase in the membrane world-volume theory, with a familiar Dirac's magnetic monopole form. Now, we are interesting in the Berry phase topic on the more complicated brane configurations. Since the Berry phase contribution is from the fermionic zero modes, we consider the path integral of the fermionic field $\chi$ such that
\begin{equation}
\int d[{\bar \chi}]d[\chi]exp[i\int d\tau{\bar \chi}(\partial_{\tau}-\gamma_a X^a)\chi ],
\end{equation}
here, the coupling of the fields $\chi$ and $X^a$ has only been remained since we want to investigate the dynamical effect of $X^a$ derived by this coupling. 

By means of the method that the Berry phase gives quantum mechanically rise to Wess-Zumino term\cite{Stone}, we can make the calculation of the effective Lagrangian part from the integration of $\chi$ and ${\bar \chi}$ become that of the Berry phase in quantum mechanics. However, a non-abelian Berry phase presents here. Because the brane configurations $X^a $ is dependent of time, the Hamiltonian ${\cal H}=X^a\gamma_a$ of the quantum system is also time-dependent. By noticing the coordinates $X^a$ with $5\leq a\leq 9$ being those of the transverse space with the symmetry $SO(5)_{\perp}$, the matrices $\gamma_a$ are the five $4\times 4$ Dirac $\gamma$ matrices given by
\begin{equation}
\gamma_{i}=\left [ \begin{array} {ll}
0&i\sigma_{i} \\
-i\sigma_{i}&0 \end{array} \right ],
\gamma_{4}=\left [ \begin{array} {ll}
0&1 \\
1&0 \end{array} \right ],
\gamma_{5}=\left [ \begin{array} {ll}
1&0\\
0&-1 \end{array} \right ].
\end{equation}
The adiabatic theorem says that for slowly varying ${\cal H}(t)$ we can approximate the solution of the time-dependent Schrodinger equation
\begin{equation}
i\partial_{t}|\psi^{\alpha}>={\cal H}^{\alpha\beta}|\psi^{\beta}>
\end{equation}
in terms of eigenstates $|\psi_{l}^{(0)}>$ of the Hamiltonian 
\begin{equation}
{\cal H}(t)|\psi_{l}^{(0)}>=E(t)|\psi_{l}^{(0)}>
\end{equation}
as
\begin{equation}
|\psi_{l}>=exp\{-i\int_{0}^{t}d\tau E(\tau)\}(P exp\int_{0}^{t}d\tau iA(\tau))_{lm}|\psi_{m}^{(0)}>=exp\{-i\int_{0}^{t}d\tau E(\tau)\}(U(t))_{lm}|\psi_{m}^{(0)}>.
\end{equation}
The Berry's non-abelian connection\cite{Wilczek} $A(t)$ is determined by the following equation
\begin{equation}
iA_{lm}+<\psi_{m}^{(0)}|\frac{\partial}{\partial t}|\psi_{l}^{(0)}>=0.
\end{equation}

Now, we consider the spherical brane configurations by adding the restrict condition $X^a X^a =R_{5}^2$, i.e., the coordinates of $S^4$ defined by $X^a/R_5$. Following Hu and Zhang\cite{Hu}, we can 
parameterize the four sphere $S^{4}$ by the following coordinate system
\begin{eqnarray}
X^1/R_5 & = & \sin \theta \sin \frac{\beta}{2} \sin(\alpha - \gamma), 
\nonumber\\
X^2/R_5 & = & - \sin \theta \sin \frac{\beta}{2} \cos(\alpha - \gamma), 
\nonumber\\
X^3/R_5 & = & - \sin \theta \cos \frac{\beta}{2} \sin(\alpha + \gamma), 
\nonumber\\
X^4/R_5 & = & \sin \theta \cos \frac{\beta}{2} \cos(\alpha + \gamma), 
\nonumber\\
X^5/R_5 & = & \cos \theta,
\end{eqnarray}
where $\theta, \beta \in[0, \pi)$ and $\alpha, \gamma \in[0, 2\pi)$. Solving the time-dependent Schrodinger equation (5), one can obtain the explicit expression of the non-abelian Berry connection. Demler and Zhang\cite{Demler} had finished the calculation of such non-abelian Berry connection in the system of a $SO(5)$ spinor coupled to a $SO(5)$ unit vector $X^a/R_5=x_a$. We can use their result to discuss the properties of the spherical brane configurations here. Their result is read as
\begin{equation}
iA^{+}=\left [ \begin{array} {ll}
<\psi^{(0)}_1|d\psi^{(0)}_1>& <\psi^{(0)}_2|d\psi^{(0)}_1> \\
<\psi^{(0)}_1|d\psi^{(0)}_2>&<\psi^{(0)}_2|d\psi^{(0)}_2>\end{array} \right ]=iA_a dx^a,
\end{equation}
where $A_a$ is given by
\begin{equation}
A_{\mu}=\frac{-1}{2(1+x_5)}\eta_{\mu\nu}^{i}x_{\nu}\sigma_{i},~~~ for ~~\mu,\nu=1,\cdots,4;
~~~~A_5 =0,
\end{equation}
where $\eta_{\mu\nu}^{i}$ is the t'Hooft symbol. $A_a$ is the $SU(2)$ gauge potential of a Yang's monopole defined on $S^4$\cite{Yang}. So the Berry connection of the spherical brane configurations induced by the background of the longitudinal 5-brane is equivalent to the Yang's $SU(2)$ monopole connection. Obviously, the connection $A^{+}$ is singular at the 'south pole' $x_5=-1$. Similarly, the another connection ${\tilde A}^{+}$ singular at the 'north pole' $x_5=1$ can also be obtained, and is given by
\begin{equation}
i{\tilde A}^{+}=\left [ \begin{array} {ll}
<\psi^{(0)}_3|d\psi^{(0)}_3>& <\psi^{(0)}_4|d\psi^{(0)}_3> \\
<\psi^{(0)}_3|d\psi^{(0)}_4>&<\psi^{(0)}_4|d\psi^{(0)}_4>\end{array} \right ]=i{\tilde A}_a dx^a .
\end{equation}
Although $A^{+}$ and ${\tilde A}^{+}$ are good at $x_5=1$ and $x_5=-1$ respectively, they are ill-defined at $x_5=-1$ and $x_5=1$ respectively. In fact we are hitting here the famous problem that for a monopole field, no single vector potential exists which is singularity-free over the entire manifold $S^4$. The use of $A^{+}$ and ${\tilde A}^{+}$ was advocated by Wu and Yang\cite{Wu} as a way round the singularity problem, since one can use each in a region where it is singularity-free, and then connect the two, in a convenient overlap region, by a gauge transformation\cite{Yang,Demler}. It is difficult to discuss the quantization and dynamical problem of the effective system only by means of the formalism of the Wu-Yang section. However, it is possible to find a singularity-free Lagrangian for the monopole problem presented here. Indeed, it is given to us automatically by the Berry phase formula, as we shall now describe.

For the case of $U(1)$ Dirac's monopole, one can obtain the effective Lagrangian related with such Berry phase by using the Balachandran formalism\cite{Balachandran}. The key step is to finish the first Hopf fibration of $S^2$ to get $S^3$. The first Hopf map is a mapping from $S^3$ to $S^2$ and is related to Dirac's monopole. In the presence of a Dirac's monopole, the $U(1)$ bundle over $S^2$ is topological non-trivial. However, since $S^3$ is parallelizable, one can use first Hops map to define a non-singular vector potential due to Dirac's monopole everywhere on $S^3$, called as the first Hopf fibration. As the above mention, the background of the longitudinal 5-brane is equivalently described by an Yang's $SU(2)$ monopole at $x^a=0$. This $SU(2)$ monopole is related with the second Hopf map $S^7\rightarrow S^4$, defined by
\begin{equation}
x_a={\bar \Psi}_{\alpha}(\gamma_a)^{\alpha\beta}\Psi_{\beta}
=Q^{\dagger}\Gamma_a Q,
\end{equation}
where $\Gamma_a$ are the five $2\times 2$ quaternionic valued gamma matrices as
\begin{eqnarray}
\Gamma_{1} & = &\left [ \begin{array} {ll}
0&-{\vec i} \\
{\vec i}&0 \end{array} \right ],
\Gamma_{2}=\left [ \begin{array} {ll}
0&-{\vec j} \\
{\vec j}&0 \end{array} \right ],
\Gamma_{3}=\left [ \begin{array} {ll}
0&-{\vec k} \\
{\vec k}&0 \end{array} \right ], \nonumber\\
\Gamma_{4} & = &\left [ \begin{array} {ll}
0&1\\
1&0 \end{array} \right ],
\Gamma_{5}=\left [ \begin{array} {ll}
1&0\\
0&-1\end{array} \right ].
\end{eqnarray}
The quaternion $Q$ can be expressed as $Q=Q^{0}+Q^{1}{\vec i}+Q^{2}{\vec j}+Q^{3}{\vec k}$ with three imaginary units ${\vec i}$, ${\vec J}$ and ${\vec k}$. From these imaginary units obeying the algebraic relations, one can easily establish the correspondence between them and Pauli matrices such that $i\sigma_1\rightarrow {\vec i}$, $i\sigma_2\rightarrow {\vec j}$ and $-i\sigma_3\rightarrow {\vec k}$. Demler and Zhang \cite{Demler} had shown that the non-abelian Berry connection can be expressed by the quaternions through the second Hopf fibration of $S^4$, i.e., $S^4\rightarrow S^7$. On $S^7$, the singularity of the non-abelian Berry connection can be completely removed by a $SU(2)$ gauge transformation. Then, the singularity-free Berry connection can be read as
\begin{equation}
iA_a dx^a=\frac{1}{2}[Q^{\dagger}_{\alpha}dQ_{\alpha}-dQ^{\dagger}_{\alpha}Q_{\alpha}],
\end{equation}
where, two quaternions $Q_{\alpha}$ with $\alpha=1,2$ are used to realize the seven-sphere by means of the relation $|Q_{1}|^2 +|Q_{2}|^2 =1$.

It can be clearly seen from the above mention that if the longitudinal 5-brane is considered as the background and the spherical 5-brane as the probe brane, we can get the effective action from the non-abelian Berry connection by integrating out the modified fields $v$ and $\chi$ in the Berkooz-Douglas model. The origin of the non-trivial Berry phase is due to the contribution from the zero modes of the field $\chi$. This Berry connection is an $SU(2)$ connection describing the Yang's $SU(2)$ monopole. In terms of the relation between the Yang's $SU(2)$ monopole and the second Hopf map, one can remove the singularity of the non-abelian Berry connection by performing the second Hopf fibration $S^4\rightarrow S^7$. If one want to consider the dynamics from the effective action and the quantum property of it, it is necessary to make the second Hopf fibration of $S^4$ since only on the $S^7$ this $SU(2)$ monopole potential can be defined both locally and globally. A similar thing happens in the case of Dirac's monopole. It is well known that since $S^3$ is the group manifold of $SU(2)$, one can reformulate the Berry connection related with the Dirac's monopole in terms of a basic dynamical variable belonging to $SU(2)$. This leads to the Balachandran form for the effective Lagrangian of $CP(1)$. In the following section, it will be shown that the $S^7$ as a geometrical object is equivalently described by a bound brane of the open membrane ending on the spherical 5-brane. On the other hand, the background of longitudinal 5-brane can be effectively described by the Yang's $SU(2)$ monopole at the center of the spherical 5-brane as the probe.

\section{Quantum Hall soliton configuration as the bound configuration of open membrane and M5-branes in M theory}

\indent

In the content of the effective 10-dimensional or 11-dimensional supergravity theory a p-brane is a solution of the field equations representing a p-dimensional extended source for an abelian (p+1)-form gauge potential $A_{p+1}$ with (p+2)-form field strength $F_{p+2}$. As such, the p-brane carries a charge
\begin{equation}
Q^{e}_{p}=\int_{S^{D-p-2}}\star F_{p+2},
\end{equation}
where $\star$ is the Hodge dual in the D-dimensional spacetime and the integral is over a $(D-p-2)$-sphere encircling the brane. $Q^{e}_{p}$ measures the electrical charge carried by the $p$-dimensional extended source since the integrating of $\star F$ is used to measure the electric charge. Similarly it is natural to define the magnetic charge. It is clear that the role that $\star F$ plays in measuring the electrical charge is played by $F$ in measuring the magnetic charge. By the electric-magnetic duality between the field strengths $F_{p+2}$ and $F_{9-p}$ in the $11$-dimensional spacetime or $F_{p+2}$ and $F_{8-p}$ in the $10$ dimensions, one can easily find
\begin{equation}
\int_{S^{D-(D-p-2)}}\star\star F_{p+2}=\int_{S^{p+2}}F_{p+2}
=Q_{7-p}^m~~for~~~ 11-d,~or~~~Q_{6-p}^m~~for~~~10-d,
\end{equation} 
which means that if $F_{p+2}$ is regarded as the field strength in the theory, the $p$-brane is the electric source, and the $(7-p)$-brane the magnetic source in $11$ dimensions or the $(6-p)$-brane the magnetic source in $10$ dimensions.

In the previous section, we obtained the conclusion that there exists equivalently an Yang's $SU(2)$ monopole at the position of the longitudinal 5-brane if this 5-brane is considered as the background in the M theory. So the 5-brane is taken as the magnetic source with the charge
\begin{equation}
Q_{5}^{m}=\int_{S^{4}}F_{4} ,
\end{equation}
where the four-index field strength $F_{\mu\nu\rho\sigma}$ should be associated with an abelian totally anti-symmetric three-index gauge potential ${\cal A}_{\mu\nu\rho}$ through the local relation $F=d{\cal A}$. In fact, this potential can be realized in terms of the non-abelian connection $A$\cite{Wu1,Demler}
\begin{equation}
{\cal A}_{\mu\nu\rho}=2Tr(A_{[\mu}\partial_{\nu}A_{\rho]}-\frac{2i}{3}A_{[\mu}A_{\nu}A_{\rho]}).
\end{equation}
Furthermore, from the point of view of the Wu-Yang section, we can calculate the magnetic charge of the 5-brane by the method of discussing the monopole. The Yang's $SU(2)$ monopole is enclosed by $S^4$. $S^4$ can be divided into two hemispheres, of which one includes the north pole of $S^4$, and the other does the south pole of $S^4$. The overlapping region between the two hemispheres is connected with the non-trivial gauge transformation. Thus, one can get
\begin{equation}
Q_{5}^{m}=\oint_{S^3_n}{\cal A}-\oint_{S^3_s}{\tilde {\cal A}}=8\pi^{2}c_2,
\end{equation}
where $c_2$ is the second Chern number. 

Following Wu and Zee\cite{Wu1}, by coupling the above three-index gauge potential to a membrane one can use an integral representation of the second Hopf mapping topological invariant to describe the phase interaction between membranes. In this description, the topological current is equivalently described by $J^{\mu\nu\rho}=(i/4!)\epsilon^{\mu\nu\rho\sigma\tau\eta\xi}F_{\sigma\tau\eta\xi}$. Thus, the effective electric charge on the membrane can be calculated by the following expression
\begin{equation}
Q_{2}^{e\prime}=\int_{S^3}J=\int_{S^3}\star F_{4},
\end{equation}
where $\star$ represents the Hodge dual in $S^7$ obtained by the second Hopf fibration from the basis space $S^4$. This charge is the background charge on the membrane. 

On the other hand, the membrane should be regarded as the electric source corresponding to the four-index field strength in $11$-dimensional spacetime. Its charge is given by $Q^{e}_{2}=\int_{S^{7}}\star F_{4}$. Since there exists the nontrivial three-index gauge potential derived by the background of the 5-brane, however, this expression should be modified as
\begin{equation}
Q^{e}_{2}=\int_{S^7}[\star F_{4}+F_{4}\wedge {\cal A}],
\end{equation}
where $S^7$ is a seven-sphere surrounding the membrane, and is from the second Hopf fibration of the basis $S^4$. We now consider the membrane with a boundary since the total charge on a compact space must vanish. Contract the $S^7$ to the boundary and deform it to the product $S^4\times S^3$ so that the entire contribution to $Q^{e}_{2}$ is given by
\begin{equation}
Q^{e}_{2}=-\int_{S^3}{\cal A}\int_{S^4}F_{4}.
\end{equation}

The above result can be interpreted as the statement that the charge of the membrane can be transferred to an electric charge of a particle on the spherical 5-brane. The existence of the open membrane\cite{Strominger,Townsend} leads to that its intersections with the 5-brane are identified with charged particles living on the 5-brane. Furthermore, the seven-sphere manifold, which is obtained from the second Hopf fibration based on the space $S^4$, is geometrical object composed of the bound configuration of the open membrane and the 5-brane. Since this manifold $S^7$ is compact, the background charge on the membrane must be cancelled out by the charge provided by the intersection of the membrane with the 5-brane, i.e., $Q^{e\prime}_{2}=-Q^{e}_{2}$. As pointed by Strominger\cite{Strominger}, the intersection of the membrane with the 5-brane gives the string boundary of the membrane in the 5-brane. This string is the self-dual string of Duff and Lu\cite{Duff}, which can be seen from that the string lying in the 5-brane must carry charge related with the 3-form field strength $\int_{S^3}{\cal A}$. Furthermore, one can define a self-dual 3-form field strength in a general background, and thus find a magnetic charge of the string boundary of the membrane in the 5-brane $Q_1^{m}=\int_{S^3}{\cal A}$. Indeed, it is a self-dual string.

Recall that our considering configuration in the M theory is the spherical 5-brane configuration in the background at the center with a longitudinal 5-brane. In such configuration, the longitudinal 5-brane background can be equivalently described by the Yang's $SU(2)$ monopole localized at the center. Globally, we must employ the second Hopf fibration $S^4\rightarrow S^7$ to describe this spherical 5-brane. Dynamically, one can say that the spherical 5-brane blow up the bound configuration of the open membrane and it in the background of the three-index anti-symmetric gauge potential derived by the longitudinal 5-brane, i.e., equivalently by the Yang's $SU(2)$ monopole. The self-dual string describes the intersection between the open membrane and the spherical 5-brane. 

In fact, the longitudinal 5-brane at the center of the spherical 5-brane together with the bound configuration of the open membrane and the spherical 5-brane should be regarded entirely as a configuration in the M theory. This configuration is denoted by $M5_{c}-M2-M5_{s}$, where the subscripts $c$ and $s$ stand for the 5-branes as the source of the background and as the probe with spherical symmetry respectively. However, we can exchange the roles of the two 5-branes, i.e., regard the spherical 5-brane as the source brane and the longitudinal 5-brane as the probe brane. Then, one can see that a bound configuration is composed of the open membrane and the longitudinal 5-brane. This implies that the open membrane is intersects with both the spherical 5-brane and the longitudinal 5-brane, with which the intersections are described by the self-dual strings in the time evolution. One of the two dimension spaces of the membrane lies in the spherical 5-brane, and the other does in the longitudinal 5-brane. The extra one dimension of it describes the time. Hence, $M5_{c}-M2-M5_{s}$ is a consistency configuration in the M theory. Interpreted in the context of the type $IIA$ string theory at low energies and weak coupling, we should be to describe arbitrary open membrane in terms of the degrees of freedom of $D0$-brane in the presence of $D4$-brane, and thus these configurations should correspond to $D0$-branes blown up into the open $D2$-branes ending on the $D4$-branes in the presence of the D4-brane. That is, the configuration $M5_{c}-M2-M5_{s}$ becomes the $D$-brane configuration $D4_{c}-D2-D4_{s}$ in the picture of the type $IIA$ string. 

Upon a conformal transformation from $S^4$ to the 4-dimensional Euclidean space $R^4$, the above $SU(2)$ monopole gauge potential is transformed to the instanton solution of the $SU(2)$ Yang-Mills theory\cite{Belavin}. For the instanton number being one and the $SU(2)$ gauge group, one finds that the unit size instanton located at the origin of $R^4$ projects to a homogeneous instanton configuration on the $S^4$. In order to discuss the properties of more general brane configurations, we shall consider some higher gauge representations of the non-abelian connection, i.e., the Yang's $SU(2)$ monopole potential. Following Zhang and Hu\cite{Zhang}, one can be very easily to do this. Such representations of the non-abelian connection can be obtained from (10) by replacing $\frac{1}{2}\sigma_{i}$ with appropriate generators $I_{i}$ of the isospin $SU(2)$ algebra, which obey the relation $[I_i,I_j]=i\epsilon_{ijk}I_k$. Replacing the fundamental $SU(2)$ generators by an $(2I+1)\times (2I+1)$ representation allows us to embed this homogeneous instanton solution in an $U(2I+1)$ gauge theory. Since we are treating all the gauge fields as a fixed background here, we can use the $U(2I+1)$ gauge field to describe effectively the $SU(2)$ gauge field with the higher representation. The instanton number of the resulting gauge field can be calculated by choosing the irreducible representation\cite{Constable}. The result is
\begin{equation}
N=\frac{1}{8\pi^2}\int_{S^4}TrF^{Y}\wedge F^{Y}=\frac{1}{6}2I(2I+1)(2I+2),
\end{equation}
where $F^{Y}$ is the gauge field strength of the instanton potential obtained by the conformal transformation of the higher representation monopole potential. 

Since $2I+1$ coincident spherical D4-branes have an $U(2I+1)$ gauge symmetry, one can think the four sphere in the background of such gauge field as the object composed of the $2I+1$ coincident spherical $D4$-branes with a homogeneous instanton of $U(2I+1)$ in their world volume\cite{Fabinger}. However, this system can be described either in terms of $D4$-branes with a $D0$-brane charge, or in terms of $D0$-branes in the presence of the $D4$-brane expanded into the bound geometrical object of one open $D2$-brane ending on the $D4$-brane, which is fuzzy due to the Myers' effect\cite{Myers}, and is described by the seven-sphere obtained by the second Hopf fibration based on $S^4$. 

If we use the latter view, and make the $N=\frac{1}{6}2I(2I+1)(2I+2)$ $D0$-branes dissolved in the bound geometrical object $S^7$, the lowest energy state of the fluctuations on the bound brane $M2-M5$ derived by the area preserving diffeomorphisms on the open membrane (see the section four) should correspond to the lowest Landau level state of the 4-dimensional quantum Hall system. This lowest energy state should be $\frac{1}{6}2I(2I+1)(2I+2)$ fold degenerate. However, the lowest Landau level state of Zhang and Hu's considering the 4-dimensional quantum Hall system is $\frac{1}{6}(2I+1)(2I+2)(2I+3)$ fold degenerate. This implies that only if $N$ ( or $I$) is enough large, the degeneracy of the former agrees with that of the latter. Hence, we call the brane configuration $M5_c -M2-M5_s$, which the $N$ $D0$-branes are spread in the bound brane $M2-M5_s$, as the 4-dimensional quantum Hall soliton configuration in the M theory. It should be pointed that one can use the seven-sphere $S^7$ to discuss the rigid property of the open membrane in the bound brane $M2-M5$ only if $N$ ( or $I$) is enough large. For this case, the quantum Hall soliton brane configuration can be stable. When $N$ is finite, we should regularize the model for the quantum Hall soliton brane configuration by adding some impurity fields to it to make the configuration become stable ( see section five in detail ).

\section{Rigid open membrane and non-commutative non-abelian Chern-Simons theory for the fluctuations on the bound brane}

\indent

In this section, we shall investigate the matrix theory describing the fluctuations of the bound brane $M2-M5$ induced by the transformations of the area preserving diffeomorphisms for the open membrane.

Since our considering the bound brane configuration $M2-M5$ can be geometrically described by the $S^7$, it is necessary to clarify the transformation properties of the coordinates ${\bf Q}_1$ and ${\bf Q}_2$ of the $S^7$ under the gauge transformations. In order to do this, we shall first recall some useful properties of the second Hopf mapping $S^7\rightarrow S^4$. Let $V$ be an element in $SU(2)$ group. $V$ can be parameterized as
\begin{equation}
V=\left [ \begin{array} {ll}
x_4+ix_3& x_2-ix_1\\
-x_2-ix_1& x_4-ix_3\end{array} \right ]
={\bf Q}_1{\bf Q}_2^{\dagger},
\end{equation}
where the first equality gives one mapping from the $SU(2)$ group to $S^3$, and the second equality decides the mapping from the $SU(2)$ group to the space of two quaternions ${\bf H}_2$. Following Hu and Zhang\cite{Hu}, one can define $(R,R^{\prime})$ to be a pair of elements in $SU(2)$, which creates the following rotation on $SU(2)$ group, $V^{\prime}=R^{-1}VR^{\prime}$. The whole set of pairs $(R,R^{\prime})$ forms a $SO(4)$ group defined in terms of the above operations. Due to the seven-sphere is characterized the normal unit condition of two quaternions $|{\bf Q}_1|^2 +|{\bf Q}_2|^2 =1$, the $SO(4)$ group operating on the seven-sphere is induced by
\begin{equation}
V\rightarrow V^{\prime}=R^{-1}{\bf Q}_1{\bf Q}_2^{\dagger}R^{\prime},
\end{equation}
and $(V^{\prime})^{\dagger}=R^{\prime -1}{\bf Q}_2{\bf Q}_1^{\dagger}R$.

It can be known from the above relations that the $SU(2)$ operations $R$ and $R^{\prime}$ of the $SO(4)$ group act on the quaternions ${\bf Q}_1$ and ${\bf Q}_2$, respectively. Based on the second Hopf mapping relation (12), we find the $SO(4)$ operations on the spinor $\Psi=(Q_1^0-iQ_1^3, Q_1^2-iQ_1^1, Q_2^0-iQ_2^3, Q_2^2-iQ_2^1)^{T}$ of $SO(5)$ as following
\begin{equation}
{\cal R}\Psi=
\left [ \begin{array} {ll}
R& {\bf 0}\\
{\bf 0}& R^{\prime}\end{array} \right ]
\left [ \begin{array} {l}
Q_1^0-iQ_1^3\\
Q_1^2-iQ_1^1\\
Q_2^0-iQ_2^3\\
Q_2^2-iQ_2^1\end{array} \right ].
\end{equation}
Here we use the quaternions ${\bf Q}_1$ and ${\bf Q}_2$ of the space ${\bf H}_2$, or the equivalent spinor $\Psi$ of $SO(5)$ defined on the complex space ${\bf C}^4$ to parameterize the seven-sphere $S^7$. In fact, this $S^7$ is the Hopf bundle obtained by the second Hopf fibration from the basis $S^4$.

As pointed by us in the section two, the background of the longitudinal 5-brane at the center can be effectively described by the Yang's $SU(2)$ monopole in the origin. Hence, this Hopf fibration can be finished by means of the relation between the second Hopf map and Yang's $SU(2)$ monopole. Geometrically, a known fact is that $\frac{SO(5)}{SU(2)\times SU(2)}=\frac{S^7}{SU(2)=S^3}=S^4$ since $S^7=\frac{SO(5)}{SU(2)}$. This implies that the $S^4$ can be identified with the coset space of the group $SO(5)$ under the $SU(2)\times SU(2)$ gauge structure. However, if one takes straightforwardly the orbital part of the $SO(5)$ generators as the generators of $SO(4)$ and decomposes them into two $SU(2)$ algebras, he can not use  such $SO(4)$ blocks to generate all irreducible representations of $SO(5)$\cite{Yang1}. Only if these orbital $SO(4)$ generators are modified by the coupling of them with a $SU(2)$ isospin, they can be used to generate the $SO(4)$ block states of all $SO(5)$ irreducible representations. Furthermore, this orbital $SO(4)$ Lie algebra is valued in the $SO(4)$ gauge fields.

Exactly, the $SU(2)$ transformations $R$ and $R^{\prime}$ are generated by two $SU(2)$ algebras of the $SO(4)$, of which one is given by a $SU(2)$ sub-algebra of the unmodified $SO(4)$ algebra, and another is produced by the coupling of the former $SU(2)$ algebra with the $SU(2)$ algebra of isospin. The local transformations associated with $R$ and $R^{\prime}$ can be characterized by the $SO(4)$ gauge field, which can be decomposed into the direct sum form of two $SU(2)$ algebras. Such $SU(2)$ gauge field is given by the $2\times 2$ matrix as the following form
\begin{equation}
{\bf A}=
\left [ \begin{array} {ll}
A_{11}& A_{12}\\
A_{21}& A_{22}\end{array} \right ]
\end{equation}
which is anti-hermitian. From (26), we can read off the property of the gauge field acting on the quaternions
\begin{equation}
\left [ \begin{array} {l}
Q_1^{0\prime}-iQ_1^{3\prime}\\
Q_1^{2\prime}-iQ_1^{1\prime}\end{array} \right ]
=\left [ \begin{array} {l}
Q_1^{0}-iQ_1^{3}\\
Q_1^{2}-iQ_1^{1}\end{array} \right ]+
\left [ \begin{array} {ll}
A_{11}& A_{12}\\
A_{21}& A_{22}\end{array} \right ]
\left [ \begin{array} {l}
Q_1^{0}-iQ_1^{3}\\
Q_1^{2}-iQ_1^{1}\end{array} \right ].
\end{equation}
By means of the conjugation of (28) and the anti-hermitian property of the matrix ${\bf A}$, the transformation relation (28) can be equivalently re-expressed as
\begin{equation}
\left [ \begin{array} {l}
Q_1^{0\prime}\\
Q_1^{3\prime}\\
Q_1^{2\prime}\\
Q_1^{1\prime}\end{array} \right ]=
\left [ \begin{array} {l}
Q_1^{0}\\
Q_1^{3}\\
Q_1^{2}\\
Q_1^{1}\end{array} \right ]+
\left [ \begin{array} {llll}
0&-\frac{i}{2}(A_{11}-A_{11}^{\ast})&
\frac{1}{2}(A_{12}+A_{12}^{\ast})&-\frac{1}{2}(A_{12}-A_{12}^{\ast})\\
\frac{i}{2}(A_{11}-A_{11}^{\ast})&0&
\frac{i}{2}(A_{12}-A_{12}^{\ast})&\frac{1}{2}(A_{12}+A_{12}^{\ast})\\
\frac{1}{2}(A_{21}+A_{21}^{\ast})&-\frac{i}{2}(A_{21}-A_{21}^{\ast})&
0&-\frac{i}{2}(A_{22}-A_{22}^{\ast})\\
\frac{i}{2}(A_{21}-A_{21}^{\ast})&\frac{1}{2}(A_{21}+A_{21}^{\ast})&
\frac{i}{2}(A_{22}-A_{22}^{\ast})&0\end{array} \right ]
\left [ \begin{array} {l}
Q_1^{0}\\
Q_1^{3}\\
Q_1^{2}\\
Q_1^{1}\end{array} \right ].
\end{equation}
Obviously, the $SU(2)$ gauge field given by (29) is an anti-symmetric matrix. If we use the indices $i$ and ${\tilde i}$ to denote the $0,3$ components and the $2,1$ components of the quaternions respectively, the matrix elements of the gauge field can be represented by $A^{ab}$ composed of $A^{ij}$, $A^{i{\tilde j}}=-A^{{\tilde j}i}$ and $A^{{\tilde i}{\tilde j}}$. This implies that under the gauge transformation of $SU(2)$, the components of the quaternions are transferred into
\begin{eqnarray}
&  Q_1^i & \rightarrow Q_1^{i\prime}=Q_1^i +A^{ij}Q_1^j+A^{i{\tilde j}}Q_1^{\tilde j},\nonumber\\
& Q_1^{\tilde i} & \rightarrow Q_1^{{\tilde i}\prime}=Q_1^{\tilde i} +A^{{\tilde i}j}Q_1^j+A^{{\tilde i}{\tilde j}}Q_1^{\tilde j}.
\end{eqnarray}
The transformation formula of the quaternions ${\bf Q}_2$ can be obtained by the replacement of $Q_1$ and $A$ in the ${\bf Q}_1$'s formula with $Q_2$ and ${\tilde A}$ respectively.

Subsequently, we shall investigate the theory describing the fluctuations of the bound configuration of the open membrane ending on the spherical $M5$-brane. As mentioned by us in the previous sections, in the language of the type IIA string theory, the open $D2$-brane ending on the $D4$-brane with the $D0$-brane charge $N$ can be thought of as $N$ $D0$-branes expanded into a spherical configuration of the above bound brane which has an induced local D4-brane charge. The expanded configuration should be fuzzy due to Myers effect for the branes. In order to investigate this fuzzy configuration, we add $N$ $D0$-branes dissolved in the primary $D2-D4$ bound configuration, i.e., the brane configuration $M2-M5_{s}$ characterized by the seven-sphere $S^7$. Since the quaternions ${\bf Q}_1$ and ${\bf Q}_2$ are used to describe this seven-sphere, they are the coordinates of the $D0$-brane on $S^7$. Furthermore, the $D0$-brane can be considered as the particle on the four-sphere $S^4$ through the second Hopf mapping $S^7\rightarrow S^4$. 

The presence of the longitudinal 5-brane at the origin of the $S^4$ induces a $D0$-brane magnetic field. From the discussions in the second section, we know that this magnetic field is given by the four-form field strength. It can be expressed locally as $H=dA$ with the components $H_{\mu\nu\lambda\rho}=H\epsilon_{\mu\nu\lambda\rho}$ and $A_{\mu\nu\lambda}=\frac{H}{4}\epsilon_{\mu\nu\lambda\rho}X^{\rho}$ where $X_{\rho}$ represent the coordinates on $S^7$. The constant field strength $H$ involves the contributions of the inverse of the $S^4$'s volume and the magnetic charge of the $M5$-brane. Because of the gauge symmetry of $SO(4)$, the three-form potential $A_{\mu\nu\lambda}$ can be chosen as the decoupling form for ${\bf Q}_1$ and ${\bf Q}_2$ such that
\begin{equation}
A_{\mu\nu\lambda}=
\{ \begin{array}{l}
\frac{H}{4}\epsilon_{a_1b_1c_1d_1}Q_1^{d_1}\\
\frac{H}{4}\epsilon_{a_2b_2c_2d_2}Q_2^{d_2} 
\end{array} . 
\end{equation}
In terms of the $SO(5)$ spinor $\Psi$, the non-abelian Berry connection (9) and (10) can be equivalently expressed \cite{Zhang} as ${\bar \Psi}d\Psi={\bar u}iA_a dx^a u$, where $u$ is an $SU(2)$ spinor. It can be seen from this that ${\bar \Psi}_i$ can be regarded as the conjugate momentum of $\Psi_i$ appearing in the effective Lagrangian. This guides us to make the further gauge choice of $A_{\mu\nu\lambda}$ consistency with the second Hopf mapping. Noticing $\Psi_1=Q_1^0-iQ_1^3$, ${\bar \Psi}_1=Q_1^0+iQ_1^3$ and $\Psi_2=Q_1^2-iQ_1^1$, ${\bar \Psi}_2=Q_1^2+iQ_1^1$, one can find that there exists only the combination of the quaternionic components $Q_1^i$ in the variable $\Psi_1$ and its conjugate momentum ${\bar \Psi}_1$ or that of $Q_1^{\tilde i}$ in the variables $\Psi_2$ and ${\bar \Psi}_2$. So the further gauge choice of $A_{\mu\nu\lambda}$ can be taken as $A_{i_1}=\frac{H}{2}\epsilon_{i_1j_1}Q_1^{j_1}$, $A_{{\tilde i}_1}=\frac{H}{2}\epsilon_{{\tilde i}_1{\tilde j}_1}Q_1^{{\tilde j}_1}$ and 
$A_{i_2}=\frac{H}{2}\epsilon_{i_2j_2}Q_2^{j_2}$, $A_{{\tilde i}_2}=\frac{H}{2}\epsilon_{{\tilde i}_2{\tilde j}_2}Q_2^{{\tilde j}_2}$ which from the similar properties of $Q_2^{i_2}$ and $Q_2^{{\tilde i}_2}$ in the $\Psi_3$ and $\Psi_4$. We call this gauge choice as taking Landau gauge of the three-form potential based on the second Hopf map. 

A $D0$-brane in the above magnetic field will experience a Lorentz force like the usual particle. The following term
\begin{equation}
L=\frac{H}{2}\epsilon_{i_1j_1}Q_1^{i_1}D_tQ_1^{j_1}+
\frac{H}{2}\epsilon_{{\tilde i}_1{\tilde j}_1}Q_1^{{\tilde i}_1}D_tQ_1^{{\tilde j}_1}+
\frac{H}{2}\epsilon_{i_2j_2}Q_2^{i_2}D_tQ_2^{j_2}+
\frac{H}{2}\epsilon_{{\tilde i}_2{\tilde j}_2}Q_2^{{\tilde i}_2}D_tQ_2^{{\tilde j}_2}
\end{equation}
will be present in its Lagrangian. The covariant derivatives are defined by $D_t Q_1^i ={\dot Q}_1^i -i[A_0^{ij},Q_1^j ]-i[A_0^{i{\tilde j}},Q_1^{\tilde j} ]$ and $D_t Q_1^{\tilde i} ={\dot Q}_1^{\tilde i} -i[A_0^{{\tilde i}j},Q_1^j ]-i[A_0^{{\tilde i}{\tilde j}},Q_1^{\tilde j} ]$. The covariant derivative for ${\bf Q}_2$ can be get by replacing $Q_1$ and $A$ with $Q_2$ and ${\tilde A}$ in the above definitions. Here we consider the low energy dynamics of the theory with the D0-brane in the case of the magnetic field strength enough strong. So the Lagrangian (32) governs the low energy dynamics of this system.

For studying the many $D0$-brane system we use the matrix theory\cite{Banks,Bernevig}. The matrix Lagrangian of the system with the $N$ $D0$-branes corresponding to (32) is read as
\begin{equation}
L=\frac{H}{2}\epsilon_{i_1j_1}Tr[Q_1^{i_1}D_t Q_1^{j_1}]+
\frac{H}{2}\epsilon_{{\tilde i}_1{\tilde j}_1}Tr[Q_1^{{\tilde i}_1}D_t Q_1^{{\tilde j}_1}]
+(2\rightarrow 1),
\end{equation}
where all dynamical variables are the $N\times N$ matrices. It can be easily shown that the Lagrangian (33) is invariant under the infinitesimal gauge transformation
\begin{eqnarray}
Q_i^{j_i}\rightarrow Q_i^{j_i} & + & i[\alpha_i^{j_i j_i^{\prime}},Q_i^{j_i^{\prime}}]
+ i[\alpha_i^{j_i {\tilde j}_i^{\prime}},Q_i^{{\tilde j}_i^{\prime}}],\nonumber\\
Q_i^{{\tilde j}_i}\rightarrow Q_i^{{\tilde j}_i}& + & i[
\alpha_i^{{\tilde j}_i j_i^{\prime}},Q_i^{j_i^{\prime}}]
+ i[\alpha_i^{{\tilde j}_i {\tilde j}_i^{\prime}},Q_i^{{\tilde j}_i^{\prime}}],\nonumber\\
A_{0(i)}^{ab}\rightarrow A_{0(i)}^{ab}& + & {\dot \alpha}_i^{ab}-i[\alpha_i^{ac},A_{0(i)}^{cb}],
\end{eqnarray}
where $i=1,2$ and $a,b,c=j,{\tilde j}$. This symmetry of the Lagrangian is a non-abelian gauge symmetry. In fact, $A_{(i)}^{ab}$ can be regarded as not only the $SO(4)$ gauge field of the target space of D0-brane $S^7$ corresponding to the bound brane configuration of the open membrane ending on the spherical 5-brane, but also the induced gauge field for area preserving diffeomorphisms of the world volume of the open membrane embedded in the target space $S^7$.

The open membrane with the invariant property of area preserving diffeomorphisms is called as the rigid open membrane by Berkooz\cite{Berkooz}. We shall consider its inducing gauge field as fluctuations of the open membrane around the classical solution of the branes, which is related with fluctuations of the bound brane configuration $M_2-M5$, to investigate the effective Lagrangian for the fluctuations. If we take $\sigma^r , r=1,2$ as the space coordinates of the open membrane embedded in $S^7$, the transformations of area preserving diffeomorphisms on the open membrane are given by
\begin{equation}
\sigma^r \rightarrow \sigma^r +\beta^r (\sigma),~~~\partial_r (w(\sigma)\beta^r (\sigma))=0,
\end{equation}
where $\beta^r $ can be locally written as
\begin{equation}
\beta^r (\sigma)=\frac{\epsilon^{rs}}{w(\sigma)}\partial_s \beta(\sigma),
\end{equation}
and $w(\sigma)$ is a 2-dimensional measure for the normalization.
The transformation rules of the fields are determined by introducing the Poisson brackets defined with respect to the measure $w(\sigma)$ as the following
\begin{equation}
\{A,B\}=\frac{\epsilon^{rs}}{w(\sigma)}\partial_r A\partial_s B.
\end{equation}

In a manner similar to the construction of the matrix theory of $D0$-branes\cite{Banks}, we replace the classical configuration space of the $N$ $D0$-branes by a space of the $N\times N$ matrices $Q_i^{a_i b_i}$. The time components of the vector potential are also replaced by the hermitian matrices. Thus, the Poisson brackets should be replaced by the commutators between the classical matrices. That is, $-i[A,B]\rightarrow \{A,B\}$. So the transformations of area preserving diffeomorphisms on the open membrane induce obviously the gauge transformations of the bound brane configuration $S^7$ given by
\begin{eqnarray}
Q_1^{i}\rightarrow  Q_1^{i} & - & i[A_1^{ij},Q_1^{j}]-i[A_1^{i{\tilde j}},Q_1^{\tilde j}]\nonumber\\
& = &
Q_1^{i} + \frac{\theta\epsilon^{rs}}{w(\sigma)}A_r^{ij}\partial_s Q_1^{j}+\frac{\theta\epsilon^{rs}}{w(\sigma)}A_r^{i{\tilde j}}\partial_s Q_1^{\tilde j},\nonumber\\
Q_1^{\tilde i} \rightarrow Q_1^{\tilde i} & - & i[A_1^{{\tilde i}j},Q_1^{j}]-i[A_1^{{\tilde i}{\tilde j}},Q_1^{\tilde j}]\nonumber\\
& = &
Q_1^{\tilde i} + \frac{\theta\epsilon^{rs}}{w(\sigma)}A_r^{{\tilde i}j}\partial_s Q_1^{j}+\frac{\theta\epsilon^{rs}}{w(\sigma)}A_r^{{\tilde i}{\tilde j}}\partial_s Q_1^{\tilde j}.
\end{eqnarray}
There exist also the induced transformation relations for the ${\bf Q}_2$ similar to the above transformations. In (38), we have introduced the deformation parameter $\theta$ characterizing the deformed size for the Myers effect of the branes, which can be fixed by the physical parameters in the system. From the definition of the Poisson brackets (37), the commutators of the coordinate matrices of $S^7$ can be read as
\begin{equation}
[ Q_i^{j_i} , Q_i^{k_i}] =  i\frac{\theta\epsilon^{rs}}{w(\sigma)}\partial_r Q_i^{j_i}\partial_s Q_i^{k_i}, ~~~
\left [ Q^{{\tilde j}_i}_i , Q^{{\tilde k}_i}_i \right ] = 
i\frac{\theta\epsilon^{rs}}{w(\sigma)}\partial_r Q_i^{\tilde j}\partial_s Q_i^{\tilde k}.
\end{equation}
It can be obviously seen from the above commutators that if the jacobian of the coordinate transformations $\partial(Q_i^{j_i},Q_i^{k_i})/\partial(\sigma^1,\sigma^2)$ is universal for the different $i$, the commutators become $[Q_i^{j_i},Q_i^{k_i}]=i\theta\epsilon^{j_i k_i}$ and $[Q_i^{{\tilde j}_i},Q_i^{{\tilde k}_i}]=i\theta\epsilon^{{\tilde j}_i {\tilde k}_i}$ by taking the suitable normalization of the measure $w(\sigma)$.

The transformation relations of ${\bf Q}_1$ and ${\bf Q}_2$ (38) can be understood as the matrix variables expanded with the fluctuations ${\bf A}$ and ${\bf {\tilde A}}$ around the classical solutions ${\bf Q}_1^{(0)}$ and ${\bf Q}_2^{(0)}$. That is, the degrees of freedom on the brane configuration $M2-M5$ arise by expanding the classical solution such that
\begin{eqnarray}
Q_1^i & = & Q_1^{i(0)}+\theta\frac{\epsilon^{rs}}{w(\sigma)}A_r^{ij}\partial_s Q_1^{j(0)}
+\theta\frac{\epsilon^{rs}}{w(\sigma)}A_r^{i{\tilde j}}\partial_s Q_1^{{\tilde j}(0)},\nonumber\\
Q_1^{\tilde i} & = & Q_1^{{\tilde i}(0)}+\theta\frac{\epsilon^{rs}}{w(\sigma)}A_r^{{\tilde i}j}\partial_s Q_1^{j(0)}
+\theta\frac{\epsilon^{rs}}{w(\sigma)}A_r^{{\tilde i}{\tilde j}}\partial_s Q_1^{{\tilde j}(0)}.
\end{eqnarray}
The expanding expressions of ${\bf Q}_2$ can be read off from (40) by replacing ${\bf A}$ with ${\bf {\tilde A}}$.

Inserting the equation (40) into the Lagrangian (33), we can find an effective Lagrangian for the fluctuations ${\bf A}$ and ${\bf {\tilde A}}$, describing the dynamics of the degrees of freedom on the bound brane configuration in the background of the classical solution. In the procedure of simplifying the effective Lagrangian, one can use the techniques of simplifying an effective Lagrangian to obtain the $U(1)$ non-commutative Chern-Simons action in \cite{Bernevig}. However, since the gauge field is non-abelian here we must use repeatedly the anti-symmetric property of $A^{ab}$, i.e, $A^{ij}=-A^{ji}$, $A^{i{\tilde j}}=-A^{{\tilde j}i}$ and $A^{{\tilde i}{\tilde j}}=-A^{{\tilde j}{\tilde i}}$. The ${\bf Q}_1$ part and the ${\bf Q}_2$ part in the Lagrangian (33) are decoupled. So the effective Lagrangian for the ${\bf Q}_2$ can be obtained by substituting the $SU(2)$ gauge field in the effective Lagrangian for the ${\bf Q}_1$ with the another $SU(2)$ gauge field. Although the calculation of simplifying the effective Lagrangian is tedious, it is straightforward. The result of the simplifying effective Lagrangian of the ${\bf Q}_1$ part is
\begin{eqnarray}
L_1^{eff} &=&\frac{H}{2}\{\epsilon_{ij}Q_1^{i}D_tQ_1^{j}+
\epsilon_{{\tilde i}{\tilde j}}Q_1^{\tilde i}D_tQ_1^{\tilde j} \}\nonumber\\
&=&\frac{H}{2}\{ Tr[i[Q_1^{i(0)},Q_1^{j(0)}]A_0^{ij}
+i [Q_1^{{\tilde i}(0)},Q_1^{{\tilde j}(0)}]A_0^{{\tilde i}{\tilde j}}\nonumber\\
&+&\frac{\theta^2}{w(\sigma)}Tr\epsilon^{\mu\nu\lambda}[A_{\mu}^{ij^{\prime}}\partial_{\nu}
A_{\lambda}^{j^{\prime}j}\epsilon^{rs}\partial_s Q_1^{j(0)}\partial_r Q_1^{i(0)}+
A_{\mu}^{i{\tilde j}^{\prime}}\partial_{\nu}
A_{\lambda}^{{\tilde j}^{\prime}j}\epsilon^{rs}\partial_s Q_1^{j(0)}\partial_r Q_1^{i(0)}\nonumber\\
& + & A_{\mu}^{{\tilde i}j^{\prime}}\partial_{\nu}
A_{\lambda}^{j^{\prime}{\tilde j}}\epsilon^{rs}\partial_s Q_1^{{\tilde j}(0)}\partial_r Q_1^{{\tilde i}(0)}+
A_{\mu}^{{\tilde i}{\tilde j}^{\prime}}\partial_{\nu}
A_{\lambda}^{{\tilde j}^{\prime}{\tilde j}}\epsilon^{rs}\partial_s Q_1^{{\tilde j}(0)}\partial_r Q_1^{{\tilde i}(0)}]\nonumber\\
&-&
\frac{2i\theta^2}{3w(\sigma)}Tr\epsilon^{\mu\nu\lambda}[A_{\mu}^{ij^{\prime}}A_{\nu}^{j^{\prime}j^{\prime\prime}}A_{\lambda}^{j^{\prime\prime}j}\epsilon^{rs}\partial_s Q_1^{j(0)}\partial_r Q_1^{i(0)}
+ A_{\mu}^{i{\tilde j}^{\prime}}A_{\nu}^{{\tilde j}^{\prime}j^{\prime\prime}}A_{\lambda}^{j^{\prime\prime}j}\epsilon^{rs}\partial_s Q_1^{j(0)}\partial_r Q_1^{i(0)}\nonumber\\
&+&
A_{\mu}^{{\tilde i}j^{\prime}}A_{\nu}^{j^{\prime}{\tilde j}^{\prime\prime}}A_{\lambda}^{{\tilde j}^{\prime\prime}{\tilde j}}\epsilon^{rs}\partial_s Q_1^{{\tilde j}(0)}\partial_r Q_1^{{\tilde i}(0)}+
A_{\mu}^{{\tilde i}{\tilde j}^{\prime}}A_{\nu}^{{\tilde j}^{\prime}{\tilde j}^{\prime\prime}}A_{\lambda}^{{\tilde j}^{\prime\prime}{\tilde j}}\epsilon^{rs}\partial_s Q_1^{{\tilde j}(0)}\partial_r Q_1^{{\tilde i}(0)}]\}\nonumber\\
& = &
\frac{H}{2}Tr \{ i[Q_1^{a(0)},Q_1^{b(0)}]A_0^{ab}+\theta[\epsilon^{\mu\nu\lambda}
A_{\mu}^{ab^{\prime}}\partial_{\nu}A_{\lambda}^{b^{\prime}b}i[Q_1^{b(0)},Q_1^{a(0)}]
\nonumber\\
&-&
\frac{2i}{3}\epsilon^{\mu\nu\lambda}A_{\mu}^{aa^{\prime}}A_{\nu}^{a^{\prime}b^{\prime}}
A_{\lambda}^{b^{\prime}b}i[Q_1^{b(0)},Q_1^{a(0)}]] \}.
\end{eqnarray}
The fundamental commutators of the classical matrices are defined by the matrix
\begin{eqnarray}
([Q_1^{a(0)},Q_1^{b(0)}]) & = &
\left [ \begin{array}{ll}
\left [Q_1^{i(0)},Q_1^{j(0)}\right]&\left [ Q_1^{i(0)},Q_1^{{\tilde j}(0)}\right ] \\
\left [Q_1^{{\tilde i}(0)},Q_1^{j(0)}\right ]& \left [ Q_1^{{\tilde i}(0)}, Q_1^{{\tilde j}(0)}\right ]
\end{array} \right ] \nonumber\\
& = &
\left [ \begin{array}{ll}
i\frac{\theta}{w(\sigma)}\epsilon^{rs}\partial_r Q_1^{i(0)}\partial_s Q_1^{j(0)}&0\\
0 &i\frac{\theta}{w(\sigma)}\epsilon^{rs}\partial_r Q_1^{{\tilde i}(0)}\partial_s Q_1^{{\tilde j}(0)}
\end{array} \right ].
\end{eqnarray}
The elements of this matrix are related with the jacobian of the variable replacement between the variables ${\bf Q}_1$ and $\sigma^r, r=1,2$ in the decoupling case with respect to the indices $i$ and ${\tilde i}$. We denote it over the measure $w(\sigma)$ as $M(\frac{\partial{\bf Q}_1}{\partial{\vec \sigma}})$.

The ${\bf Q}_1^{(0)}$ and ${\bf Q}_2^{(0)}$ are the matrices of the classical solution to be identified with the non-commutative coordinates of the bound brane $M2-M5$.
Since any matrix can be expressed in terms of finite sum of products $\prod_{ab}exp\{ip_aQ_1^{a(0)}\}exp\{ip_bQ_2^{b(0)}\}$, the $N\times N$ matrices $A_{\mu}^{ab}$ can be thought of as functions of the ${\bf Q}_1^{(0)}$ and ${\bf Q}_2^{(0)}$. Based on this fact, we can make the effective Lagrangian pass to the continuum limit taking $N$ large. The changes of the coordinates $\sigma^r, r=1,2$ parameterizing the open membrane induce the variations of the bound brane coordinates ${\bf Q}_1^{(0)}$ and ${\bf Q}_2^{(0)}$. Furthermore, the fields $A_{\mu}^{ab}$ and ${\tilde A}_{\mu}^{ab}$ are the degrees of freedom describing the fluctuations on the open membrane. So we can identify the trace of matrix $\theta Tr$ as $\int\frac{d\sigma^1 d\sigma^2}{2\pi}$. In the continuum limit, the $N\times N$ matrices $A_{\mu}^{ab}$ will map to smooth functions of the non-commutative coordinates ${\bf Q}_1^{(0)}$ and ${\bf Q}_2^{(0)}$. For the field as the functions of non-commutative coordinates, we can introduce the Weyl ordering to define a suitable ordering for their products in the effective Lagrangian. This implies that the ordinary product should be replaced by the non-commutative $\star$-product. Finishing all of these, we find the effective action corresponding to the Lagrangian (41)
\begin{eqnarray}
S[A] & = &\frac{H}{4\pi}\int d^3\sigma[(-\Sigma^{ab})M(\frac{\partial{\bf Q}_1}{\partial {\vec \sigma}})\star A_0^{ab}+\theta\epsilon^{\mu\nu\lambda}M(\frac{\partial{\bf Q}_1}{\partial {\vec \sigma}})\star (A_{\mu}^{ab}\star \partial_{\nu}A_{\lambda}^{ba}-\frac{2i}{3}A_{\mu}^{ab}\star A_{\nu}^{bc}\star A_{\lambda}^{ca})]\nonumber\\
& = &\frac{H}{4\pi}\int d^3\sigma[(-\Sigma^{ab})M(\frac{\partial{\bf Q}_1}{\partial {\vec \sigma}})\star A_0^{ab}+\theta\epsilon^{\mu\nu\lambda}M(\frac{\partial{\bf Q}_1}{\partial {\vec \sigma}})\star tr(A_{\mu}\star \partial_{\nu}A_{\lambda}-\frac{2i}{3}A_{\mu}\star A_{\nu}\star A_{\lambda})],
\end{eqnarray}
where $\Sigma^{ab}=\epsilon^{ij}$ for $a,b=i,j$, $\epsilon^{{\tilde i}{\tilde j}}$ for $a,b={\tilde i},{\tilde j}$ and is vanishing otherwise. The first term in the action is the chemical potential term for the ${\bf Q}_1$ part provided by the background charge on the open membrane in the presence of the longitudinal 5-brane. The jacobian $M(\frac{\partial{\bf Q}_1}{\partial {\vec \sigma}})$ exhibits the characteristic property that the non-commutativity on the open membrane is from the non-commutativity of the bound brane coordinates ${\bf Q}_1$ and ${\bf Q}_2$, which are parameterized by the world volume variables $(t,\sigma^1,\sigma^2)$. The second term in it is a $SU(2)$ non-commutative Chern-Simons action. The level of this Chern-Simons theory is given by $H\theta$. 

As the previous mention, the total effective action of the system is the sum of the ${\bf Q}_1$ part and the ${\bf Q}_2$ part since they are decoupled through the $SO(4)$ gauge symmetry. The effective action for the ${\bf Q}_2$ can be obtained by finishing the discussion paralleling to that for the ${\bf Q}_1$ part. Only difference is the replacement of the $SU(2)$ gauge field $A_{\mu}^{ab}$ with the another $SU(2)$ gauge field ${\tilde A}_{\mu}^{ab}$. So we get the total effective action for the fluctuations around the classical solution being
\begin{equation}
S=S[A]+S[{\tilde A}].
\end{equation}

Up to now, we have given the effective description for the fluctuations. Our conclusion is that the dynamics of the fluctuations on the bound brane derived by the area preserving diffeomorphisms of the open membrane is described by the $SO(4)$ non-commutative Chern-Simons theory.

\section{The non-commutative structure of $S^4$ and the finite matrix non-commutative Chern-Simons theory}

\indent

First of all, we shall use the non-commutative structure of the $S^7$ derived by the rigid open membrane to exhibit the non-commutativity of the $S^4$ by means of the second Hopf mapping. Let us recall some useful properties of the second Hopf mapping\cite{Demler,Zhang}. From the definition (12) of the second Hopf map and the fundamental properties of the quaternions, we have the following mapping relations
\begin{eqnarray}
x_1 & = & 2(Q_1^0 Q_2^1 -Q_1^1 Q_2^0 +Q_1^2 Q_2^3 -Q_1^3 Q_2^2 ),\nonumber\\
x_2 & = & 2(Q_1^0 Q_2^2 -Q_1^2 Q_2^0 +Q_1^3 Q_2^1 -Q_1^1 Q_2^3 ),\nonumber\\  
x_3 & = & 2(Q_1^0 Q_2^3 -Q_1^3 Q_2^0 +Q_1^1 Q_2^2 -Q_1^2 Q_2^1 ),\nonumber\\ 
x_4 & = & 2(Q_1^0 Q_2^0 +Q_1^1 Q_2^1 +Q_1^2 Q_2^2 +Q_1^3 Q_2^3 ),\nonumber\\ 
x_5 & = & (Q_1^0 )^2 -(Q_2^0 )^2 +(Q_1^1 )^2 -(Q_2^1 )^2 +(Q_1^2 )^2 -(Q_2^2 )^2+(Q_1^3 )^2 -(Q_2^3 )^2. 
\end{eqnarray}

In order to simply exhibit the non-commutativities of the brane configurations, we consider the system with a $D0$-brane on the $S^7$ in the presence of the longitudinal 5-brane at the origin. Since the $S^7$ should be regarded as the bound configuration of the open membrane ending on the spherical 5-brane, there exist some non-trivial Poisson brackets for the coordinates of the D0-brane on the $S^7$ in the background of the longitudinal 5-brane. These non-trivial Poisson brackets generate the gauge transformations of the fundamental fields ${\bf Q}_1$ and ${\bf Q}_2$ induced by the transformations of area preserving diffeomorphisms for the open membrane. Based on the discussion in the above section, we know that they are given by
\begin{eqnarray}
\{Q_1^i ,Q_1^j \} & = & \theta\frac{\epsilon^{rs}}{w(\sigma)}\partial_r Q_1^i \partial_s Q_1^j ,~~~\{Q_1^{\tilde i} ,Q_1^{\tilde j} \}=\theta\frac{\epsilon^{rs}}{w(\sigma)}\partial_r Q_1^{\tilde i} \partial_s Q_1^{\tilde j} , \nonumber\\
\{Q_2^i ,Q_2^j \} & = &\theta\frac{\epsilon^{rs}}{w(\sigma)}\partial_r Q_2^i \partial_s Q_2^j ,~~~\{Q_2^{\tilde i} ,Q_2^{\tilde j} \}=\theta\frac{\epsilon^{rs}}{w(\sigma)}\partial_r Q_2^{\tilde i} \partial_s Q_2^{\tilde j}.
\end{eqnarray}
In fact, the term $\epsilon^{rs}\partial_r Q_1^i \partial_s Q_1^j$ can be expressed as $\epsilon^{ij}M(\frac{\partial(Q_1^0 ,Q_1^3 )}{\partial(\sigma^1 ,\sigma^2 )})$, i.e., the product of the anti-symmetric tensor with the jacobian of the variable replacement between $Q_1^0 ,Q_1^3 $ and $\sigma^1 ,\sigma^2 $. The jacobians for ${\bf Q}_1 $ and ${\bf Q}_2$ with respect to $\sigma^1 ,\sigma^2 $ can be chosen to be uniform, which means that $M(\frac{\partial(Q_l^i ,Q_l^j )}{\partial(\sigma^1 ,\sigma^2 )})=M(\frac{\partial(Q_l^{\tilde i} ,Q_l^{\tilde j} )}{\partial(\sigma^1 ,\sigma^2 )})=M$ for $l=1,2$. Furthermore, we let $w(\sigma)$ be equal to $M$ for the normalization. Thus, the Poisson algebraic relations (46) become
\begin{eqnarray}
\{Q_1^i ,Q_1^j \} & = & \theta\epsilon^{ij},~~~~\{Q_1^{\tilde i} ,Q_1^{\tilde j} \}=\theta\epsilon^{{\tilde i}{\tilde j}},\nonumber\\
\{Q_2^i ,Q_2^j \} & = & \theta\epsilon^{ij},~~~~\{Q_2^{\tilde i} ,Q_2^{\tilde j} \}=\theta\epsilon^{{\tilde i}{\tilde j}}.
\end{eqnarray}

By straightforwardly calculating in terms of the relations (45) and (47), we have found the coordinates on $S^4$ obeying the Poisson algebraic relations such that
\begin{eqnarray}
\{x_1,x_2\} & = & 4\theta( (Q_1^1 )^2 +(Q_1^2 )^2-(Q_1^0 )^2 -(Q_1^3 )^2\nonumber\\
& + & (Q_2^1 )^2 +(Q_2^2 )^2 -(Q_2^0 )^2  -(Q_2^3 )^2)=4\theta{\bar \Psi}
\left [ \begin{array}{ll}
-\sigma_3 &0\\
0&\sigma_3
\end{array} \right ]\Psi,\nonumber\\
\{x_1,x_3\} & = & 8\theta(Q_1^2 Q_1^3 +Q_2^2 Q_2^3-Q_1^1 Q_1^0  -Q_2^1 Q_2^0 )=
4\theta{\bar \Psi}
\left [ \begin{array}{ll}
\sigma_2 &0\\
0&\sigma_2
\end{array} \right ]\Psi,\nonumber\\
\{x_1,x_4\} & = & 8\theta(Q_2^1 Q_2^3 +Q_2^2 Q_2^0 -Q_1^1 Q_1^3 -Q_1^2 Q_1^0 )=
4\theta{\bar \Psi}
\left [ \begin{array}{ll}
-\sigma_1 &0\\
0&\sigma_1
\end{array} \right ]\Psi,\nonumber\\
\{x_1,x_5\} & = & 8\theta(Q_2^1 Q_1^3 +Q_2^2 Q_1^0 +Q_1^1 Q_2^3 +Q_1^2 Q_2^0 )=
4\theta{\bar \Psi}
\left [ \begin{array}{ll}
0&\sigma_1\\
\sigma_1&0
\end{array} \right ]\Psi,\nonumber\\
\{x_2,x_3\} & = & 8\theta(Q_2^1 Q_2^3 +Q_2^2 Q_2^0 +Q_1^1 Q_1^3 +Q_1^2 Q_1^0 )=
4\theta{\bar \Psi}
\left [ \begin{array}{ll}
\sigma_1 &0\\
0&\sigma_1
\end{array} \right ]\Psi,\nonumber\\
\{x_2,x_4\} & = & 8\theta(Q_1^1 Q_1^0 +Q_2^2 Q_2^3 -Q_1^2 Q_1^3 -Q_2^1 Q_2^0 )=
4\theta{\bar \Psi}
\left [ \begin{array}{ll}
-\sigma_2 &0\\
0&\sigma_2
\end{array} \right ]\Psi,\nonumber\\
\{x_2,x_5\} & = & 8\theta(Q_2^2 Q_1^3 -Q_2^1 Q_1^0 +Q_1^2 Q_2^3 -Q_1^1 Q_2^0 )=
4\theta{\bar \Psi}
\left [ \begin{array}{ll}
0&\sigma_2\\
\sigma_2&0
\end{array} \right ]\Psi,\nonumber\\
\{x_3,x_4\} & = & 4\theta( (Q_1^1 )^2 +(Q_1^2 )^2-(Q_1^0 )^2 -(Q_1^3 )^2\nonumber\\
& + & (Q_2^0 )^2  +(Q_2^3 )^2 -(Q_2^1 )^2 -(Q_2^2 )^2 )=4\theta{\bar \Psi}
\left [ \begin{array}{ll}
\sigma_3 &0\\
0&\sigma_3
\end{array} \right ]\Psi,\nonumber\\
\{x_3,x_5\} & = & 8\theta(Q_2^3 Q_1^3 +Q_2^0 Q_1^0 -Q_1^1 Q_2^1 -Q_1^2 Q_2^2 )=
4\theta{\bar \Psi}
\left [ \begin{array}{ll}
0&\sigma_3\\
\sigma_3&0
\end{array} \right ]\Psi,\nonumber\\
\{x_4,x_5\} & = & 8\theta(Q_2^0 Q_1^3 -Q_2^3 Q_1^0 +Q_1^1 Q_2^2 -Q_1^2 Q_2^1 )=
4\theta{\bar \Psi}
\left [ \begin{array}{ll}
0&-i1\\
i1&0
\end{array} \right ]\Psi.
\end{eqnarray}
The second equalities in (48) can be easily checked by means of the relation $(\Psi)^T=(Q_1^0 -iQ_1^3 ,Q_1^2 -iQ_1^1 , Q_2^0 -iQ_2^3 ,Q_2^2 -iQ_2^1 )$ of the $SO(5)$
spinor $\Psi$ with the quaternions. 

Since an $SO(5)$ spinor can be represented by a four component complex vector $\Psi_{\alpha}$, each spinor of $SO(5)$ can be mapped into an $SO(5)$ vector, i.e., the unit vector normal to the $S^4$. Zhang and Hu \cite{Zhang} had found the solution of this mapping. Explicitly, the orbital coordinate $x_a$, which is defined by the coordinate point of the 4-dimensional sphere $X_a = R x_a$, is related with the spinor coordinates $\Psi_\alpha$ with $\alpha=1, 2, 3, 4$ by the relations $x_a = \bar{\Psi} \gamma_a \Psi$ 
and $\sum_{\alpha} \bar{\Psi}_\alpha \Psi_\alpha = 1$. The isospin 
coordinates ${\bf n}_i = \bar{u}{\bf \sigma}_i u$ with $i=1, 2, 3$ are 
given by an arbitrary two-component complex spinor $(u_1, u_2)$ satisfying 
$\sum_{\sigma} \bar{u}_\sigma u_\sigma = 1$. Zhang and Hu gave the 
explicit solution of the spinor coordinate with respect to the orbital 
coordinate as following
\begin{equation}
\left ( \begin{array} {l}
\Psi_1 \\
\Psi_2 \end{array} \right ) = \sqrt{\frac{1 + x_5}{2}} \left ( 
\begin{array} {l}
u_1 \\
u_2 \end{array} \right ),\quad
\left ( \begin{array} {l}
\Psi_3 \\
\Psi_4 \end{array} \right ) = \sqrt{\frac{1}{2(1+x_5)}} (x_4 - i x_i 
\sigma_i) \left ( \begin{array} {l}
u_1 \\
u_2 \end{array} \right ).
\end{equation}

Substituting the solution (49) of the $SO(5)$ spinor into the Poisson algebraic relations (48), we find that the Poisson brackets between the coordinates on the $S^4$ can be rewritten as the following compact form
\begin{equation}
\{x_a,x_b\}=4\theta{\bar u}f_{ab}^i \sigma_i u=4\theta f_{ab}^i n_i ,
\end{equation}
where $f_{ab}^i $ is the gauge field strength of the Yang's $SU(2)$ monopole, given by $f_{\mu\nu}^i =x_{\nu}a_{\mu}^i -x_{\mu}a_{\nu}^i \eta_{\mu\nu}^i $ and $f_{5\mu}^i =-(1+x_5 )a_{\mu}^i $ for $\mu,\nu=1,2,3,4$. This result is agree with the conclusion in \cite{Zhang,Zhang1}. It should be emphasized that although the Poisson brackets of the coordinates on the $S^4$ is related with the $SU(2)$ gauge field strength valued in the $SU(2)$ algebra of the isospin, this $SU(2)$ algebra must be projected out through the inner product of it between the spinor and its conjugation of the isospin. Only if one does this Poisson algebra is closed. In other words, one must extended the space $S^4$ to the seven-dimensional space including the three-dimensional isospin space to make the Poisson algebra on the $S^4$ closed. However, such seven-dimensional space is the space of the bound brane configuration $M_2-M_5$ in which the $D0$-brane living. The longitudinal 5-brane at the origin provided a constant magnetic field strength as the background. This makes the coordinates of the bound configuration with a non-commutative structure. The non-commutativity of the coordinates on the spherical 5-brane is just induced by the non-commutative structure of the bound brane configuration through the second Hopf mapping. On the other hand, the unit vector ${\vec n}$ can be realized by the higher dimensional representation of the $SU(2)$ algebra, i.e., by $n_i ={\bar U}_m (I_i  )_{mm^\prime}U_{m^\prime}/I(I+1)$ for $m,m^{\prime}=I,I-1,\cdots,-I$. Here $I_i$ are the generators of $2I+1$ dimensional representations of the $SU(2)$ Lie algebra. Hence, the above conclusion is also right for an arbitrary representation of the $SU(2)$ Lie algebra of isospin. Our result shows that the non-commutative relations (47) of the coordinates on the $S^7$ are more fundamental, and they induce the non-commutativity of the coordinates on the $S^4$.

For the case of many $D0$-branes, we can use the $N\times N$ matrices $x_a$ and ${\bf Q}_1$, ${\bf Q}_2$ to describe their coordinates on the $S^4$ and on the $S^7$ respectively. However, the matrices $x_a$ are the functions of ${\bf Q}_1$ and ${\bf Q}_2$ obeying the non-trivial algebraic relations of the classical matrices. If we make them pass to the continuum limit taking $N$ large, $x_a$ are mapped to the functions of the non-commutative coordinates ${\bf Q}_1$ and ${\bf Q}_2$. As mentioned above, we should introduce the Weyl ordering to define the ordering for the products of the components of ${\bf Q}_1$ and ${\bf Q}_2$. So now we must use the non-commutative $\star$-product to replace the ordinary product in the Poisson brackets of $x_a$. Therefore, from the Poisson relation (50), we read off the Poisson algebraic relation of the coordinates $x_a$ on the $S^4$ for many $D0$-branes as following
\begin{equation}  
\{x_a\star ,x_b\}=4\theta f_{ab}^i(x,\star) n_i ,
\end{equation}

We come now to the description of the fluctuations of the rigid open membrane with the $N$ $D0$-branes, where $N$ is finite. The motivations for this description have two fold. One of them is to discuss the relation between the 4-dimensional quantum Hall system and the brane configuration $M5_c -M_2 -M5_s$ in M theory. This brane configuration is possibly stable in the limit of taking $N$ infinite. But one must consider the quantum Hall system with the finite number of particles to establish the microscopic description of the quantum Hall effect. Taking the number of particles infinite means that the thermodynamic limit of this quantum system is taken. The other is to use the fluctuation configurations of the rigid open membrane to interpret the blow-up of the ADHM moduli space of instantons\cite{Berkooz}. If one considers the finite dimensions of the moduli space, he needs to give the description of the fluctuations of the rigid open membrane with the finite $N$ $D0$-branes.

The description of the fluctuations of the rigid open membrane with the finite $N$ $D0$-branes can be from the finite $N$ regularization of the non-abelian non-commutative Chern-Simons theory given by us in the section four. Polychronakos\cite{Polychronakos} had proposed a matrix regularized version of the $U(1)$ non-commutative Chern-Simons theory proposed by Susskind\cite{Susskind}. Here we shall give the non-abelian generalization of this matrix regularized version to the non-abelian non-commutative Chern-Simons theory. The action of the generally regularized matrix model is read as
\begin{equation}
S_{M} =S[Z^{\alpha}, \Psi^{\alpha}, A_{0}]+S[Z^{\bar \alpha}, \Psi^{\bar \alpha}, {\bar A}_{0}] . 
\end{equation}
This matrix model are divided into two $SU(2)$ parts according to the $SO(4)$ gauge symmetry of the non-commutative Chern-Simons theory (44). The explicit form of action for the $SU(2)$ part is given by 
\begin{eqnarray}
S[Z^{\alpha}, \Psi^{\alpha}, A_{0}] & = & \frac{H}{4}\int dt Tr\{ {\bar Z}^{\alpha} i(\partial_t Z^{\alpha} +[A_{0}^{\alpha\beta},Z^{\beta}])- 2i\theta\sigma_{3}A_{0}\nonumber\\ 
& - & \omega {\bar Z}^{\alpha}Z^{\alpha} \}
+\frac{1}{2}\int dt\Psi^{\dagger\alpha}i(\partial_t \Psi^{\alpha} +A_{0}^{\alpha\beta}\Psi^{\beta})+ h.c.
\end{eqnarray}
Here, we consider the system composed of $N$ $N0$-branes. The complex coordinates $Z^{\alpha}$, which can be expressed as 
$Z^{1}=Q_1^0 +iQ_1^3 $ and $Z^{2}=Q_1^1 +iQ_1^2 $, are represented by finite $N\times N$ matrices. The 
$2\times 2$ matrix built up by the gauge fields $A_{0}^{\alpha\beta}$ with the indices $(\alpha,\beta)$  is anti-hermitian and 
traceless except for that each element of them is a $N\times N$ matrix.  
$\theta$ is the positive parameter characterizing the non-commutativity of the coordinates of $D0$-branes, and $\sigma_{3}$ the third component of Pauli matrix. The another $S[Z^{{\bar \alpha}}, \Psi^{{\bar \alpha}}, {\bar A}_{0}]$ of two $SU(2)$ parts in (52) can be obtained from (53) by the replacement of $Z^{\alpha}$ with $Z^{{\bar \alpha}}$, which is defined by $Z^{{\bar 1}}=Z^{3}=Q^0_2+iQ^3_2 , Z^{{\bar 2}}=Z^{4}=Q^1_2+iQ^2_2$.

Each element of $\Psi^{\alpha}$ and $\Psi^{\bar \alpha}$ is a complex $N$-vector that transforms in the fundamental of the gauge group $U(N)$. Their parts in the action are the covariant kinetic terms similar to the complex scalar fermions. However, we shall quantize them as the bosons. Since there are no spatial kinetic terms for them that would lead to the negative Dirac seas, it is perfectly consistency to quantize them as bosons. They can be also interpreted as the impure fields like as Berkooz's explanation for the open membrane in the six-dimensional (2,0) field theory. If one would like to keep the topology of a closed sphere even when discussing the world sheet of open membrane, he can make the closed sphere into open sphere by inserting an 'impurity' at some point along the sphere.

Because of our considering the fluctuations on the bound brane $M2-M5$ derived by the area preserving diffeomorphisms on the open membrane, the effective matrix model (52) should provide the dynamics of fluctuations on the spherical 5-brane, i.e., $S^4$, by means of the second Hopf mapping $S^7\rightarrow S^4$. On the $S^4$, one cannot entirely establish the Lagrangian formalism for the fluctuations due to the topologically non-trivial property of the second Hopf map. Hence, we should add the geometrically restricted condition for the second Hopf map to the model (52) to describe the physics on the $S^4$. 
Two pairs of complex coordinates $Z^{\alpha}$ and $Z^{{\bar \alpha}}$ of the complex space ${\bf C}^{4}$ can be expressed as $Z^{1}=Q^0_1+iQ^3_1 , Z^{2}=Q^1_1+iQ^2_1$ and $Z^{{\bar 1}}=Z^{3}=Q^0_2+iQ^3_2 , Z^{{\bar 2}}=Z^{4}=Q^1_2+iQ^2_2$
in terms of eight real coordinates of the space ${\bf H}^{2}$ of quaternions equivalent to ${\bf R}^{8}$.  We can get the restricted condition of $S^{7}$ by considering the sphere
in ${\bf R}^8$, which is $ |{\bf Q}_{1}|^2 +|{\bf Q}_{2}|^2 ={\bar Z}^{a}Z^{a} =1$. Now $Z^{a}$ are finite $N\times N$ matrices. Therefore, we write the geometrically restricted condition of the model as
\begin{equation}
Tr({\bar Z}^{a}Z^{a}) =G ,
\end{equation}
where $G$ is a parameter dependent on the model. This implies that our matrix regularized model described by the action (52) together with the geometrical restricted condition (54) of $S^{7}$ which is from the second Hopf fibration of $S^{4}$.

The equations of motion of $A_{0}$ and ${\bar A}_{0}$
are taken as the constraint equations
\begin{eqnarray}
\frac{H}{2}[Z^{1},{\bar Z}^{1}] & - & \frac{H}{2}[Z^{2},{\bar Z}^{2}]+\Psi^{1}\Psi^{1\dagger}-\Psi^{2}\Psi^{2\dagger}=2H\theta, \nonumber\\
\frac{H}{2}[Z^{2},{\bar Z}^{1}] & + & \Psi^{2}\Psi^{1\dagger}=0. \nonumber\\
\frac{H}{2}[Z^{3},{\bar Z}^{3}] & - & \frac{H}{2}[Z^{4},{\bar Z}^{4}]+\Psi^{3}\Psi^{3\dagger}-\Psi^{4}\Psi^{4\dagger}=2H\theta, \nonumber\\
\frac{H}{2}[Z^{4},{\bar Z}^{3}] & + & \Psi^{4}\Psi^{3\dagger}=0.
\end{eqnarray}
If one ignores the contributions from the impure fields, he can easily find that the commutators of the classical matrices $[Q_r^i ,Q_r^j ]=i\theta\epsilon^{ij}$ and $[Q_r^{\tilde i} ,Q_r^{\tilde j} ]=i\theta\epsilon^{{\tilde i}{\tilde j}}$ for $r=1,2$ are the solution of the equations (55). In fact, in the limit of taking $N$ infinite, the contributions from the fields $\Psi^{\alpha}$ and $\Psi^{\bar \alpha}$ can be ignored. So passing to the continuum limit taking $N$ infinite, the constraint equations of our matrix model lead indeed reduce to the non-commutativity of the non-abelian Chern-Simons theory (44). Conclusively, the matrix model given by us here is just the regularized version of the $SO(4)$ non-commutative Chern-Simons theory. It is the starting point to study the 4-dimensional quantum Hall fluids in \cite{Chen1,Chen2}.

\section{Summary and outlook}

\indent

Similar to how the 2-dimensional quantum Hall system is modeled by the string interacting with the $D$-branes, we showed how the 4-dimensional quantum Hall system of Zhang and Hu can be modeled by the open membrane interacting with the $M5$-branes in M theory, or by the open $D2$-brane doing with the $D4$-branes in the language of the type $IIA$ string theory. In this formalism, the longitudinal 5-brane at the center plays the role of providing a constant four-form field strength for the probe brane, which is equivalently described by an Yang's $SU(2)$ monopole at the center. An open membrane is naturally ended on the 5-branes by the topological consistency of the brane configuration. The open membrane ending on the spherical 5-brane is made of the bound brane configuration which the $N$ $D0$-branes live on. In the limit taking the $N$ as infinite, the low energy dynamics of the quantum Hall soliton brane configuration $M5_c -M2-M5_s $ can effectively describe the essential physics of the 4-dimensional quantum Hall system. If the $N$ is finite, we should regularize the model of the quantum Hall soliton brane configuration. In order to use the spherical geometrical object $S^7$ to describe the bound brane configuration $M2-M5_s $ in the presence of the four-form field strength, we must introduce some impure fields in the regularized version of the model. The reason of doing this is from the fact that a similar thing happens in string theory. One can make the closed sphere into an open sphere by inserting an impurity at some point along the sphere. This impurity will allow the fields to be non-smooth around this point, which will in effect make it into an open disk. Obviously, the quantum Hall soliton brane configuration constructed by us here is different from the string realization of the Zhang and Hu's 4-dimensional quantum Hall effect proposed by Fabinger\cite{Fabinger}. His realization is provided by that some fundamental strings are stretched between the spherical $D4$-brane and a stack of flat $D4$-branes placed at the center of the $S^4$. It is interesting to clarify the relation between our quantum Hall soliton brane configuration and Fabinger's string realization.

The rigid property of the open membrane plays the dominant role in the higher dimensional quantum Hall soliton brane construction here. Justly, the fluctuations of the $D0$-branes around the classical solution induced by the transformations of area preserving diffeomorphisms on the open membrane, i.e., the rigid open membrane, provide the low energy dynamical degrees of freedom describing the excitations of the quantum Hall soliton brane configuration. The effective field theory for them is the $SO(4)$ non-commutative Chern-Simons field theory given by us here. However, the open membrane cannot be isolated from the bound brane configuration $M2-M5_s $ because of the topologically non-trivial property of the second Hopf fibration $S^4\rightarrow S^7$. Hence, the bound brane configuration $M2-M5_s $ must be parameterized by embedding the open membrane into this bound configuration. If one investigates the physics on the spherical 5-brane $S^4$, he should perform the second Hopf mapping $S^7\rightarrow S^4$. Based on this fact, we found the non-commuative structure of the $S^4$ from the non-commutativity of the coordinates on the bound brane $M2_M5_s $ by finishing this second Hopf mapping. We produced the result in agreement with Zhang and Hu's that in \cite{Zhang,Zhang1}. So the non-commutative structure of the coordinates on the bound brane is more fundamental.

In order to make the quantum Hall soliton brane configuration to become stable with the finite $D0$-branes, we have regularized the $SO(4)$ non-commutative field theory. Our finding the matrix regularized model is the non-abelian generalization of the matrix Chern-Simons model for the 2-dimensional quantum Hall effect proposed by Polychronakos\cite{Polychronakos}. Such matrix regularized model had been used for us to investigate the second quantized descriptions of the 4-dimensional quantum Hall fluids\cite{Chen1,Chen2}. In \cite{Chen1,Chen2}, we have determined the physical quantum states and the consistent hierarchical structures of the 4-dimensional quantum Hall fluids by solving the constraint equations of the physical states. 

On the other hand, it should be noticed that the constraint equation (55) of the matrix regularized model (52) has the similar form as the motion of equation in the six-dimensional (2,0) field theory. The approach in this paper can be investigate the six-dimensional (2,0) field theory\cite{Aharony,Aharony1} by exchanging the roles of the longitudinal 5-brane and the spherical 5-brane. It is well known that there does not exist any the effective Lagrange formalism of the six-dimensional (2,0) field theory up to now. In fact, such local formalism does not exist. However, it is possible to establish the effectively non-commutative Lagrange formalism since the non-commutative field theory is non-local itself. In order to do this, we should make the two sets of the constraint equations in (55) reduce one set by transferring the four-form field strength into the $U(1)$ quantum number of one $SU(2)$ and preserving the model invariant under the other symmetries. We believe that this should provide the effective Lagrange formalism for the six-dimensional (2,0) field theory, and it is useful to exhibit the properties of the physical states and the correlation functions. Furthermore, the constraint equations (55) is with the structure in the agreement with the equations of non-commutative instantons\cite{Konechny}. The approach in this paper may be used to describe the quantum mechanics on the deformed moduli space of the instantons by the resolution of the instanton singularity. The related topics as mentioned above are in progress. \\

I would like to thank Bo-Yu Hou for many valuable conversations. 
The work was partly supported by the NNSF of China (Grant No.90203003) and by the Foundation 
of Education Ministry of China (Grant No.010335025).


\end{document}